\documentclass[10pt,journal,compsoc]{IEEEtran}
\usepackage[pdftex]{graphicx,xcolor}
\usepackage{url}

\usepackage[fleqn]{amsmath} 
\usepackage{newtxtext} 
\usepackage[varg]{newtxmath} 

\usepackage{latexsym}
\usepackage{verbatim}
\usepackage{subfigure}
\usepackage[all]{xy}
\usepackage[none]{hyphenat}

\usepackage{bbding}
\usepackage{pifont}
\usepackage{wasysym}
\usepackage{amssymb}

\newif\iftodo
\todotrue

\newif\iftable
\tabletrue

\newif\ifmemo
\memofalse

\newif\ifmark
\markfalse

\newif\ifrelatedwork
\relatedworkfalse

\newtheorem{theorem}{Theorem}
\newtheorem{assumption}{Assumption}

\newcommand\tablength{0.3}
\newcommand\figwidth{80mm}

\newcommand\htab{\hspace{\tablength cm}}
\newcommand\encolor{purple}

\ifCLASSOPTIONcompsoc
  \usepackage[nocompress]{cite}
\else
  \usepackage{cite}
\fi
\ifCLASSINFOpdf
\else
\fi
\begin{document}
%
\title{Cost- and Energy-Aware Multi-Flow Mobile Data Offloading Using Markov Decision Process}
%
%
%
%
\author{Cheng~Zhang$^1$,~
        Bo~Gu$^2$,~
        Zhi~Liu$^3$,~
        Kyoko~YAMORI$^4$,~
        and~Yoshiaki~TANAKA$^5$\\
        \footnotesize{$^1$Department of Computer Science and Communications Engineering,
Waseda University, Tokyo, 169-0072 Japan,Email: cheng.zhang@akane.waseda.jp\\
       $^2$Department of Information and Communications Engineering,
Kogakuin University, Tokyo, 192-0015 Japan\\
       $^3$Department of Mathematical and Systems Engineering, Shizuoka University, Shizuoka, 432-8561 Japan\\
        $^4$Department of Management Information, Asahi University, Mizuho-shi, 501-0296 Japan\\
        $^5$Department of Communications and Computer Engineering,
Waseda University, Tokyo, 169-8555 Japan}

}

\IEEEtitleabstractindextext{%
\begin{abstract}
With 
\ifmark
\textcolor{\encolor}{the}
\else
the
\fi
rapid increase in demand for mobile data,
mobile network operators are trying to expand wireless network capacity
by deploying wireless local area network (LAN) 
\ifmark
\textcolor{\encolor}{hotspots on which they can offload their mobile traffic.}
\else
hotspots on which they can offload their mobile traffic.
\fi
However, these network-centric methods
usually do not 
\ifmark
\textcolor{\encolor}{fulfill the interests of mobile users (MUs). }
\else
fulfill the interests of mobile users (MUs). 
\fi
Taking into consideration many issues, 
\ifmark
\textcolor{\encolor}{MUs should be able to decide whether to offload their traffic}
\else
MUs should be able to decide whether to offload their traffic
\fi
to a complementary wireless LAN. Our previous 
\ifmark
\textcolor{\encolor}{work studied single-flow wireless LAN offloading from}
\else
work studied single-flow wireless LAN offloading from
\fi
a MU's perspective by considering delay-tolerance of traffic, 
monetary cost and energy consumption.
In this paper, 
\ifmark
\textcolor{blue}{we study the multi-flow mobile data offloading problem from a MU's perspective}
\else
we study the multi-flow mobile data offloading problem from a MU's perspective
\fi
in which a MU has multiple applications to download data simultaneously 
from remote servers, and different applications' data have different deadlines. 
We formulate the wireless LAN offloading problem as a
finite-horizon discrete-time Markov decision process (MDP) 
\ifmark
\textcolor{\encolor}{and establish an optimal policy by a}
\else
and establish an optimal policy by a
\fi
dynamic programming based algorithm.  
\ifmark
\textcolor{\encolor}{Since the time complexity of the dynamic}
\else
Since the time complexity of the dynamic
\fi
programming based offloading algorithm is still high, 
\ifmark
\textcolor{\encolor}{we propose a low}
\else
we propose a low
\fi
time complexity heuristic offloading algorithm with performance sacrifice. 
Extensive simulations are conducted to validate our proposed offloading algorithms.
\end{abstract}

}

\maketitle

\IEEEdisplaynontitleabstractindextext

%
\IEEEpeerreviewmaketitle

\section{Introduction}\label{intro}
  The mobile data traffic demand is growing rapidly. According to the investigation
  of Cisco Systems \cite{CiscoVNI2015}, the mobile data traffic is expected
  to reach 24.3 exabytes per month by 2019, while it was only 2.5 exabytes per
  month at the end of 2014.  On the other hand, the growth rate of the mobile 
  network capacity is far from satisfying that kind of the demand, 
  which has become a major problem for wireless mobile network operators
  (MNOs). Even though 5G technology is promising for providing huge wireless
  network capacity \cite{5GCapacity2014}, the development process is long and
  the cost is high. Economic methods such as time-dependent pricing
  \cite{TimeDependentPriceDuopolyNSP}\cite{TimeDependentPriceOligopolyNSP}
  have been proposed to change users' usage pattern, which are not user-friendly.
  Up to now, the best practice for increasing mobile network capacity is to deploy
  complementary networks (such as wireless LAN and femtocells), which can be quickly
  deployed and are cost-efficient. Using such methods, part of the MUs' traffic demand
  can be offloaded from a MNO's cellular network to the complementary network.\\
  \indent The process that a mobile device automatically changes its connection
  type (such as from cellular network to wireless LAN) is called \emph{vertical handover} \cite{Marquez-Barja:2011}.
  Mobile data offloading is facilitated by new standards such as Hotspot 2.0 \cite{Hotspot20}
  and the 3GPP Access Network Discovery and Selection Function (ANDSF)
  standard \cite{3GPP4WiFi}, with which information of network (such as price and
  network load) can be broadcasted to MUs in real-time. Then MUs can make offloading
  decisions intelligently based on the real-time network information.\\
  \indent There are many works related to the wireless LAN offloading problem.
  However, previous works either considered the wireless LAN offloading problem 
  from the network providers' perspective without considering the MU's 
  quality of service (QoS)
  \cite{BargainingOffloading}\cite{DoubleAuction2014}, or studied wireless LAN offloading
  from the MU's perspective \cite{Aug3GWiFi2010}\cite{LeeHowMuchWiFi2013}
  \cite{AMUSEMungChiang2013}\cite{DAWNHuang2015},
  but without taking the energy consumption as well as cost problems into consideration.\\
  \indent Our previous work \cite{ChengAPNOMS2016} studied the wireless LAN offloading
  problem from the MU's perspective. The MU's target was to minimize its total cost, while taking monetary cost, preference for energy consumption,
  availability of MU's mobility pattern and application's delay tolerance into consideration.
  A Markov decision process algorithm \cite{LiuZhiAccess}\cite{LiuZhiConfMDP2011}\cite{LiuZhiTrans2013} was proposed for a known 
  MU's mobility pattern case and a reinforcement learning \cite{RLBook} 
  based algorithm was proposed for an unknown MU's mobility pattern case. 
  However, \cite{ChengAPNOMS2016}  only
  considered a MU's single flow. Actually, a MU always execute mulitple applications
  simultaneously with modern mobile devices 
  \ifmark
  \textcolor{\encolor}{that have powerful multi-task abilities.}
  \else
  that have powerful multi-task abilities.
  \fi
  Therefore, multi-flow mobile data offloading problem from a MU's perspective 
  \ifmark
  \textcolor{\encolor}{is more relevant and remains to be solved.}\\
  \else
   is more relevant and remains to be solved.\\
  \fi
  \indent In this paper, we study the wireless LAN offloading problem from 
  a MU's perspective considering multi-flow.  Each flow has different delay 
  tolerance.  The MU's target is to minimize its total cost, which includes 
  the monetary cost and energy consumption cost, while taking the MNO's 
  usage base price, the MU's preference for energy consumption, 
  and flows' delay tolerance into consideration.
  The cost- and energy-aware wireless LAN offloading problem is modeled as a finite-horizon
  discrete-time Markov decision process (FDTMDP) under the assumption that
  the MU's mobility pattern is known in advance. We propose a dynamic programming
  based algorithm to solve the FDTMDP problem. However, the time complexity of the
  dynamic programming based offloading algorithm is high. Therefore,
  we propose a heuristic offloading algorithm with low time complexity
  and performance sacrifice. We conduct the simulations to verify the performance
  of the proposed schemes, and the simulation results show that the dynamic programming based offloading algorithm can minimize the MU's monetary cost
  and save energy of the MU's device, while the heuristic offloading algorithm
  has comparable performance in terms of cost minimization and energy saving
  for the MU.\\
\ifmark
\indent \textcolor{red}{The proposed mobile data offloading 
algorithms can be implemented on the MUs' device without 
modification of the network system.  The MUs themselves, 
or third-party application developers can utilize our work
to save monetary cost and energy for the MUs.}\\
\else
\indent The proposed mobile data offloading 
algorithms can be implemented on the MUs' device without 
modification of the network system.  The MUs themselves, 
or third-party application developers can utilize our work
to save monetary cost and energy for the MUs.\\
\fi
  \indent The rest of this paper is organized as follows.
  Section \ref{relatedwork} describes the related work.
  Section \ref{systemmodel} illustrates the system model.
  Section \ref{dpalgorithm} formulates the user's
  wireless LAN offloading problem as discrete-time finite-horizon
  Markov decision process and proposes a dynamic
  programming based algorithm.
  Section \ref{heuristicalgorithm} proposes a
  low time complexity heuristic offloading algorithm.
  Section \ref{performanceevaluation} illustrates the
  simulation and results. Finally, we conclude this
  paper in Section \ref{conclusion}.
\section{Related Work}\label{relatedwork}
  Mobile data offloading has been widely studied
  in the past. Gao et al. \cite{BargainingOffloading}
  studied the cooperation among one MNO and multiple
  access point owners (APOs) by utilizing the Nash bargaining
  theory, and the case of multiple MNOs and multiple APOs 
  is studied in \cite{DoubleAuction2014},
  where double auctions were adopted.
  The aforementioned papers \cite{BargainingOffloading}\cite{DoubleAuction2014}
  considered the mobile data offloading market from the perspective
  of the network without considering the MU's experience directly.\\
  \indent On the other hand, papers\cite{Aug3GWiFi2010}\cite{LeeHowMuchWiFi2013}\cite{AMUSEMungChiang2013}\cite{DAWNHuang2015}
  have considered offloading delay-tolerant traffic from the MUs'
  perspective. In \cite{Aug3GWiFi2010}, Balasubramanian et al.
  implemented a prototype system called \emph{Wiffler} to
  leverage delay-tolerant traffic and fast switching to 3G.
  Im et al. \cite{AMUSEMungChiang2013}
  not only took a MU's throughput-delay tradeoffs into
  account, but also considered the MU's 3G budget explicitly.
  A MU's mobility pattern was predicted by a second-order
  Markov chain. In \cite{DAWNHuang2015}, Cheung studied
  the problem of offloading delay-tolerant applications for
  each user. A Markov decision process was formulated to
  minimize total data usage payment. Similar to \cite{DAWNHuang2015}, 
  Kim et al. in \cite{MutiFlowOffloading2016} also utilized 
  a Markov decision process based approach to allocate
  cellular network or wireless LAN data rate to maximize a MU's 
  satisfaction, which only depended on the MU's wireless LAN usage.\\
  \indent The above literature does not consider the energy
  consumption problem when offloading traffic from
  a cellular network to a complementary network. Actually,
  the battery life has always been a concern for smartphones.
  \cite{EnergyOffloading3G2011}\cite{EnergyCollaborate2015}
  have studied how to design an energy-efficient framework
  for mobile data offloading. However, the trade off between
  throughput, delay and budget constraints have not been considered in
  these works. While it was shown in \cite{LeeHowMuchWiFi2013}
  that wireless LAN data offloading saved 55\% of battery power
  due to the much higher data rate wireless LAN can provide,
  it was verified in \cite{EnergyCollaborate2015}
  that wireless LAN could consume more energy than
  cellular network when wireless LAN throughput was lower.
  In order to clarify the contradiction, it is necessary to
  consider energy consumption to establish a cost- and
  energy-aware mobile data offloading scheme.\\
  \indent Our previous work \cite{ChengAPNOMS2016} studied the wireless LAN offloading
  problem from a MU's perspective. The MU's target was to minimize its total cost under
  usage based pricing, while taking monetary cost, preference for energy consumption,
  availability of the MU's mobility pattern and application's delay tolerance into consideration.
  A Markov decision process algorithm was proposed for a known MU's mobility pattern
  case and a reinforcement learning \cite{RLBook} based algorithm was proposed
  for an unknown MU's mobility pattern case. However, \cite{ChengAPNOMS2016}  only
  considered a MU's single flow case. \\
  \indent Different from the aforementioned papers, in this paper, we study
  a multi-flow mobile data offloading problem in which a MU has multiple
  applications to transmit data simultaneously with different deadlines,
  as well as considering the MU's monetary cost and energy consumption.
 \begin{figure}[t]
   \centering
   \includegraphics[width=70mm]{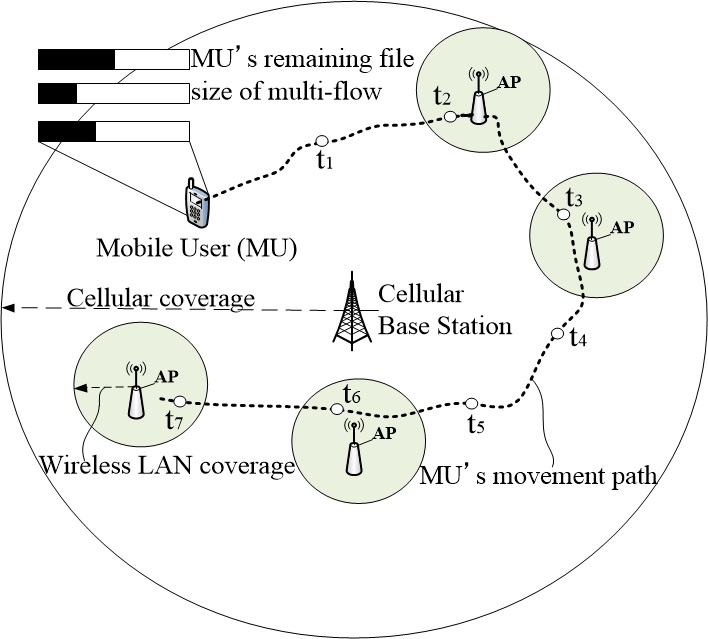}
   \caption{System scenario.}\label{scenarioFig}
 \end{figure}

\section{System Model}\label{systemmodel}
  Since the cellular network coverage is rather high, it is assumed that
  the MU is always in a cellular network, but not always can access wireless LAN
  access points (APs). The wireless LAN APs are usually deployed at home,
  stations, shopping malls and so on. Therefore, we assume that wireless LAN
  access is location-dependent (see Fig. \ref{scenarioFig}).
 \ifmark
   \textcolor{red}{We mainly focus on applications with data of relative large size and delay-tolerance to download, for example,  applications like software update, file download, email with attachments.}
 \else
	We mainly focus on applications with data of relative large size and delay-tolerance to download, for example,  applications like software updates, file downloads, or emails with attachments.
 \fi
  The MU has $M$ files to download from a remote server. 
  Each file formulates a flow, and the set of flows is denoted as
  $\mathcal{M}$$=$$\{1,...,M\}$. Each flow $j\in\mathcal{M}$ has a deadline
  $T^j$. $\textbf{\textit{T}}$$=$$(T^1, T^2, ... , T^M)$ is the 
  deadline vector for the MU's $M$ flows. Please note that we only consider
  downlink communication in this paper.
  Without loss of generality, it is
  assumed that $T^1 \le T^2 \le  ... \le T^M$. We consider a slotted time
  system as $t$$\in$$\mathcal{T}$$=$$\{1,...,T^M\}$.
  To simplify the analysis, we use limited discrete locations instead of
  infinite continous locations. It is assumed that a MU can move in $L$ 
  possible locations, which is denoted as set $\mathcal{L}$$=$$\{1,...,L\}$.
  While the cellular network is available at all the locations, the availability of 
  wireless LAN network is dependent on location $l\in\mathcal{L}$.
  The MU has to make a decision on what network to select and how to allocate the
  available data rate among $M$ flows at location $l$ at time $t$ by considering
  total monetary cost, energy consumption and remaining time for data transmission.
  A MU's mobility can be modelled by a Markovian model as in \cite{AMUSEMungChiang2013}\cite{DAWNHuang2015}.
  Therefore, the MU's decision making problem can be
  modelled as a finite-horizon Markov decision process.\\
  \begin{figure}[t]
   \centering
   \includegraphics[width=80mm]{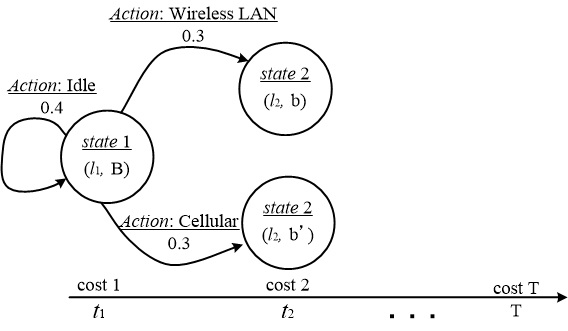}
   \caption{An example of MDP modelling: at time $t_1$, the \textit{state} 1
   contains location $l_1$ and remaining file size B.  MU chooses actions of
   \textit{Wireless LAN}, \textit{Cellular network}, or \textit{Idle}, which incur
   different cost on MU.  The fraction number under the \textit{action} is the
   transition probability which depends on MU's mobility pattern. The objective
   of MU is to 
 \ifmark
   \textcolor{blue}{minimize}
 \else
   minimize
 \fi
    total cost from time $1$ to $T$ (cost 1 + cost 2 +  ... + cost T). }\label{mdpFig}
 \end{figure}
  \indent We define the system \textit{state} at $t$ as in Eq. (\ref{state})
  \begin{equation}\label{state}
    \textbf{\textit{s}}_t = \{l_t,\textbf{\textit{b}}_t\}
  \end{equation}
  where $l_t\in \mathcal{L}$$=$$\{1,...,L\}$ is the MU's
  location index at time $t$, which can be obtained from GPS. 
  $\mathcal{L}$ is the location set. 
  $\textbf{\textit{b}}_t$$=$$(b_t^1, b_t^2, ... , b_t^M)$ is the vector of 
  remaining file sizes of all $M$ flows at time $t$, $b_t^j\in \mathcal{B}^j$
  $=[0,B^j]$ for all $j\in \mathcal{M}$.
  $B^j$ is the total remaining data size for flow $j$.
  $\mathcal{B}$=$(\mathcal{B}^1,\mathcal{B}^2, ... , \mathcal{B}^M)$,
  is the set vector of remaining data.\\
  \indent The MU's \textit{action} $a_t$ at each decision epoch $t$ 
  is to determine whether to transmit data through wireless LAN
  (if wireless LAN is available), or cellular network, or just
  keep idle and how to allocate the network data rate to $M$
  flows. Therefore, the MU's action vector is denoted as in Eq. (\ref{action})
  \begin{equation}\label{action}
      \textbf{\textit{a}}_{t}=(\textbf{\textit{a}}_{t,c},\textbf{\textit{a}}_{t,w})
  \end{equation}
  where $\textbf{\textit{a}}_{t,c}=(a_{t,c}^1, a_{t,c}^2, ..., a_{t,c}^M)$ denotes the
  vector of cellular network allocated data rates, $a_{t,c}^j$  denotes the cellular data
  rate allocated to flow $j\in$$\mathcal{M}$, and
  $\textbf{\textit{a}}_{t,w}=(a_{t,w}^1, a_{t,w}^2, ... , a_{t,w}^M)$ denotes the vector of
  wireless LAN network allocated data rates, and $a_{t,w}^j$  denotes
  the wireless LAN rate allocated to flow $j\in$$\mathcal{M}$.
  Here the subscript $c$ and $w$ stand for cellular network 
  and wireless LAN, respectively.
  Please note that $a_{t,w}^1$, $a_{t,w}^2$, ..., $a_{t,w}^M$
  all can be 0 if the MU is not in the coverage area of wireless LAN AP. 
  Even though it is technically possible that wireless LAN and cellular network can 
  be used at the same time, we 
\ifmark
  \textcolor{blue}{assume} 
\else
  assume
\fi
  that the MU can not use wireless LAN and 
  cellular network at the same time. 
\ifmark
\textcolor{red}{
  We make this assumption for two reasons:
  (i) If we restrict the MU to use only one network interface at the same time slot,
  then the MU's device may be used for longer time for the same amount of left battery.
  (ii) Nowadays smartphones, such as an iPhone, can only use one network interface 
  at the same time. We can easily implement our algorithms on a MU's device without
  changing the hardware or OS of the smartphone if we have this assumption.}
  \else
  We make this assumption for two reasons:
  (i) If we restrict the MU to use only one network interface at the same time slot,
  then the MU's device may be used for longer time for the same amount of left battery.
  (ii) Nowadays smartphones, such as an iPhone, can only use one network interface 
  at the same time. We can easily implement our algorithms on a MU's device without
  changing the hardware or OS of the smartphone if we have this assumption.
  \fi
  \ifmark
  \textcolor{blue}{At time $t$, MU may choose to use wireless LAN (if wireless LAN 
  is available) or cellular network, or not to use any network. If the MU chooses wireless 
  LAN at $t$, the wireless LAN network allocated data rate to flow
  $j$, $a_{t,w}^j$, is greater than 0, and the MU does not use cellular network 
  in this case, then $a_{t,c}^j$ = 0.  On the other hand, if the MU chooses 
  cellular network at $t$, the cellular network allocated data rate to 
  flow $j$, $a_{t,c}^j$, is greater than 0, and the MU does not 
   use wireless LAN in this case, then $a_{t,w}^j$ = 0}.
  \else
At time $t$, MU may choose to use wireless LAN (if wireless LAN 
  is available) or cellular network, or not to use any network. If the MU chooses wireless 
  LAN at $t$, the wireless LAN network allocated data rate to flow
  $j$, $a_{t,w}^j$, is greater than 0, and the MU does not use cellular network 
  in this case, then $a_{t,c}^j$ = 0.  On the other hand, if the MU chooses 
  cellular network at $t$, the cellular network allocated data rate to 
  flow $j$, $a_{t,c}^j$, is greater than 0, and the MU does not 
   use wireless LAN in this case, then $a_{t,w}^j$ = 0.
  \fi  
  $a_{t,n}^j$, $n\in\{c,w\}$ should not be greater than the remaining file 
  size $b_t^j$ for flow $j\in$$\mathcal{M}$.\\
  \indent The sum data rate of all the flows of cellular network and wireless LAN
  are denoted as $a_{t,c}=\sum_{j\in \mathcal{M}}a_{t,c}^j$ and $a_{t,w}=\sum_{j\in\mathcal{M}}a_{t,w}^j$, 
  respectively. $a_{t,c}$ and $a_{t,w}$ should satisfy the following
  conditions.
  \begin{equation}
  a_{t,c} \le \gamma_{c}^l
  \end{equation}
  \begin{equation}
  a_{t,w} \le \gamma_{w}^l
  \end{equation}
  $\gamma_{c}^l$ and $\gamma_{w}^l$ are the maximum data rates of cellular network and
  wireless LAN, respectively, at each location $l$.\\
  \iftable
\begin{table}[t]\label{table1}
\caption{Notations summary.}
\renewcommand{\arraystretch}{1}
\begin{tabular}{cp{6.3cm}}
\hline

\hline
Notation & Description\\
\hline

\hline
$\mathcal{M}$ & $\mathcal{M}$$=$$\{1,...,M\}$, MU's $M$ flows set. \\
$\textbf{\textit{T}}$ & $\textbf{\textit{T}}$$=$$(T^1, T^2, ... , T^M)$, MU's deadline vector. \\
$t$ & $t\in\mathcal{T}^M$, the specific decision epoch of MU. \\
$\mathcal{L}$ & $\mathcal{L}$$=$$\{1,...,L\}$, the location set of MU. \\
$\mathcal{B}^j$ & $\mathcal{B}^j$$\subseteq$$[0,...,b_t^j]$, the total size of MU's $j$ flow. $j\in \mathcal{M}$. \\
$\textbf{\textit{b}}_t$ & $\textbf{\textit{b}}_t$$=$$(b_t^1, b_t^2, ... , b_t^M)$, vector of remaining file size. \\
$\textbf{\textit{s}}_t$ & $\textbf{\textit{s}}_t=(l_t,\textbf{\textit{b}}_t)$, state of MU.\\
$l_t$   & $l_t\in \mathcal{L}$, MU's location index at time $t$.\\
$a_{t,c}^j$  & cellular data rate allocated to flow $j\in$$\mathcal{M}$ at time $t$\\
$a_{t,w}^j$ &  wireless LAN data rate allocated to flow $j\in$$\mathcal{M}$ at time $t$\\
$\textbf{\textit{a}}_{t,c}$&$\textbf{\textit{a}}_{t,c}=\{a_{t,c}^1, a_{t,c}^2, ..., a_{t,c}^M\}$\\
$\textbf{\textit{a}}_{t,w}$&$\textbf{\textit{a}}_{t,w}=\{a_{t,w}^1, a_{t,w}^2, ..., a_{t,w}^M\}$\\
$\textbf{\textit{a}}_{t}$ &$\textbf{\textit{a}}_{t}=(\textbf{\textit{a}}_{t,c},\textbf{\textit{a}}_{t,w})$\\
$ a_{t,c}$&$ a_{t,c}= \sum_{j\in \mathcal{M}} a_{t,c}^j$\\
$a_{t,w}$&$ a_{t,w}= \sum_{j\in \mathcal{M}} a_{t,w}^j$\\
$\gamma_{c}^l$ & cellular throughput in bps at location $l$.\\
$\gamma_{w}^l$ & wireless LAN throughput in bps at location $l$.\\
$\varepsilon_c^l$ & energy consumption rate of celllar network  in joule/bits at location $l$.\\
$\varepsilon_w^l$ & energy consumption rate of wireless LAN in joule/bits at location $l$.\\
$\theta_t$ & energy preference of MU at $t$.\\
$p_c$ & MNO's usage-based price for cellular network service.\\
$\hat{c}_{T^M+1}(\cdot)$&MU's penalty function for remaining data at $T^M+1$.\\
$\xi_t(\textbf{\textit{s}}_t,\textbf{\textit{a}}_t)$& MU's energy consumption at $t$.\\
$\phi_t$ & $\mathcal{L}$$\times$$\mathcal{K}$$\rightarrow$$\mathcal{A}$, transmission decision at $t$.\\
$\pi$ & $\pi=\{\phi_t(l,b),\;\forall\; t\in\mathcal{T^M}, l\in\mathcal{L}, b\in\mathcal{B}\}$, MU's policy.\\
\hline

\hline
\end{tabular} \\
\end{table}
\else
\fi
  \indent At each epoch $t$, three factors affect the MU's decision.
  \begin{itemize}
    \item (1) \textit{the monetary cost}: it is the payment from the MU
    to the network service provider. We assume that the network
    service provider adopts $\textit{usage-based pricing}$, which
    is being widely used by carriers in Japan, USA, etc.
    The MNO's price is denoted as $p_c$. It is assumed that
    wireless LAN is free of charge. We define the monetary cost
    $c_t(\textbf{\textit{s}}_t,\textbf{\textit{a}}_t)$ as in Eq. (\ref{payment})
    \begin{equation}\label{payment}
       c_t(\textbf{\textit{s}}_t,\textbf{\textit{a}}_t)= p_c\sum_{j\in\mathcal{M}}\min\{b_t^j,a_{t,c}^j\}
    \end{equation}
    \item (2) \textit{the energy consumption}: it is the energy consumed
    when transmitting data through wireless LAN or cellular network. We denote
    the MU's awareness of energy as in Eq. (\ref{energy})
    \begin{equation}\label{energy}
    \begin{split}
        \xi_t(\textbf{\textit{s}}_t,\textbf{\textit{a}}_{t})
        =&\theta_t (\varepsilon_c^l\sum_{j\in\mathcal{M}}\min\{b_t^j,a_{t,c}^j\}\\
        &+ \varepsilon_w^l\sum_{j\in\mathcal{M}}\min\{b_t^j,a_{t,w}^j\})
    \end{split}
    \end{equation}
    where $\varepsilon_c^l$ is the energy consumption rate of the cellular network
    in joule/bits at location $l$ and $\varepsilon_w^l$ is the energy consumption rate 
	of the wireless LAN in joule/bits at location $l$. 
    It has been shown in \cite{EnergyCollaborate2015} that 
    both $\varepsilon_c^l$ and $\varepsilon_w^l$
    decrease with throughput, which means that
    low transmission speed consumes
    more energy when transmitting the same amount of data.
  \ifmark
  \textcolor{red}{
  According to \cite{EnergyDownUplink}, the energy consumptions for downlink 
  and uplink are different. Therefore, the energy consumption parameters $\varepsilon_c^l$
  and  $\varepsilon_w^l$ should be differentiated for downlink or uplink, respectively.
  In this paper, we do not differentiate the parameters for downlink or uplink because
  only the downlink case is considered.  Nevertheless, our proposed algorithms 
  are also applicable for uplink scenarios with energy consumption parameters for uplink.}
  \else
  According to \cite{EnergyDownUplink}, the energy consumptions for downlink 
  and uplink are different. Therefore, the energy consumption parameters $\varepsilon_c^l$
  and  $\varepsilon_w^l$ should be differentiated for downlink or uplink, respectively.
  In this paper, we do not differentiate the parameters for downlink or uplink because
  only the downlink case is considered.  Nevertheless, our proposed algorithms 
  are also applicable for uplink scenarios with energy consumption parameters for uplink.
  \fi 
    $\theta_t$ is the MU's preference for energy consumption at time $t$.
    $\theta_t$ is the weight on energy consumption
    set by the MU. Small $\theta_t$ means that the MU cares less
    on energy consumption. For example, if the MU can soon charge his
    smartphone, he may set $\theta_t$ to a small value,
    or if the MU is in an urgent status and could not charge
    within a short time, he may set a large value for $\theta_t$.
    $\theta_t$ = 0 means that the MU does not consider energy consumption at all
    in the process of data offloading, just like in \cite{DAWNHuang2015} \cite{MutiFlowOffloading2016}.
    \item (3) \textit{the penalty}: if the data transmission
    does not finish in deadline $T^j$, $j\in\mathcal{M}$,
    the penalty for the MU is defined as Eq. (\ref{penalty}).
    \begin{equation}\label{penalty}
        \hat{c}_{T^j+1}(\textbf{\textit{s}}_{T^j+1})=\hat{c}_{T^j+1}(l_{T^j+1},\textbf{\textit{b}}_{T^j+1})=g(\textbf{\textit{b}}_{T^j+1})
    \end{equation}
    where $g(\cdot)$ is a non-negative non-decreasing function.
    ${T^j+1}$ means that the penalty is calculated after deadline
    $T^j$.
  \end{itemize}
  \indent The probabilities associated with different state
  changes are called transition probabilities. We denote
  \textit{transition probability} as in Eq. (\ref{transitionprob})
  \begin{equation}\label{transitionprob}
        \textrm{Pr}(\textbf{\textit{s}}_{t+1}|\textbf{\textit{s}}_t,\textbf{\textit{a}}_{t})
  \end{equation}
  Eq. (\ref{transitionprob}) shows the probability of
  state $\textbf{\textit{s}}_{t+1}$ if action $\textbf{\textit{s}}_t$ is chosen at
  state $\textbf{\textit{s}}_t$. It is assumed that the remaining
  size is independent of location change, therefore
    \begin{equation}\label{pr}
    \begin{split}
        \textrm{Pr}(\textbf{\textit{s}}_{t+1}|\textbf{\textit{s}}_t,\textbf{\textit{a}}_{t}) & = \textrm{Pr}((l_{t+1},\textbf{\textit{b}}_{t+1})|(l_t,\textbf{\textit{b}}_{t}),\textbf{\textit{a}}_{t})\\
        &= p_{l_{t+1},l_t}\textrm{Pr}(\textbf{\textit{b}}_{t+1}|(l_t,\textbf{\textit{b}}_t),\textbf{\textit{a}}_{t})
    \end{split}
    \end{equation}
  where
  \begin{equation}
  \begin{split}
    &\textrm{Pr}(\textbf{\textit{b}}_{t+1} | (l_t,\textbf{\textit{b}}_t),\textbf{\textit{a}}_{t})\\
    &= \left\{ \begin{array}{ll}
    1 & \textrm{if $\textbf{\textit{b}}_{t+1}=[\textbf{\textit{b}}_t-\textbf{\textit{a}}_{t,c}-\textbf{\textit{a}}_{t,w}]^{+}$ }\\
    0 & \textrm{otherwise}
    \end{array} \right.
  \end{split}
  \end{equation}
  $[\textbf{\textit{x}}]^{+}$ is equal to $\max\{\textbf{\textit{x}},0\}$.
  The MU's probability from $l$ to $l_{t+1}$ is denoted as $p_{l_{t+1},l_t}$, 
  which is assumed as known (see Assumption \ref{mobileassumption}).
\begin{assumption}\label{mobileassumption}
  The MU's mobile probability to move from the current location to the next
  location is known in advance.
\end{assumption}
 \indent The MU's mobility pattern can be derived from the MU's
 historical data, which has been widely studied in the
 literature, such as \cite{AMUSEMungChiang2013}.\\
 \indent The MU's \textit{policy} is defined as in Eq. (\ref{policy})
 \begin{equation}\label{policy}
   \pi=\bigg\{\phi_t(l_t,\textbf{\textit{b}}_t),\;\forall\; t\in\mathcal{T},l\in\mathcal{L}, \textbf{\textit{b}}_t\in\mathcal{B}\bigg\}
 \end{equation}
 where $\phi_t(l_t,\textbf{\textit{b}}_t)$ is a function mapping from state
 $\textbf{\textit{s}}_t=(l_t,\textbf{\textit{b}}_t)$ to a decision action at $t$.
 The set of $\pi$ is denoted as $\Pi$. If policy $\pi$
 is adopted, the state is denoted as $\textbf{\textit{s}}_t^{\pi}$.\\
 \indent The objective of the MU is to the minimize the expected
 total cost (include the monetary cost and the energy consumption)
 from $t=1$ to $t=T^M$ and penalty at $t=$${T^M+1}$ with an
 a optimal $\pi^*$ (see Eq. (\ref{totalcost}))
 \begin{equation}\label{totalcost}
    \min_{\pi\in\Pi} E_{\textbf{\textit{s}}_1}^{\pi}
    \Bigg[
    \sum_{t=1}^{T^M}r_t(\textbf{\textit{s}}_t^{\pi},\textbf{\textit{a}}_{t})
    +\sum_{j\in\mathcal{M}}\hat{c}_{T^j+1}(\textbf{\textit{s}}_{T^j+1}^{\pi})
    \Bigg]
\end{equation}
where $r_t(\textbf{\textit{s}}_t,\textbf{\textit{a}}_{t})$ is the sum of the
 monetary cost and the energy consumption as in Eq. (\ref{reward})
 \begin{equation}\label{reward}
    r_t(\textbf{\textit{s}}_t,\textbf{\textit{a}}_{t})=c_t(\textbf{\textit{s}}_t,\textbf{\textit{a}}_{t})+\xi_t(\textbf{\textit{s}}_t,\textbf{\textit{a}}_{t})
 \end{equation}
\indent Please note that the optimal action at each $t$ does not lead
to the optimal solution for the problem in Eq. (\ref{totalcost}).  At each
time $t$, not only the cost for the current time $t$ should be considered,
but also the future expected cost. \\
\indent Please refer to Fig. \ref{mdpFig} for an example of a MDP modelling
in this section. The notations used throughout this paper are summarized
as shown in Table 1.

\section{Dynamic Programming Based Algorithm}\label{dpalgorithm}
 The MU's network selection and rate allocation problem has been formulated
 as a standard finite-horizon discrete-time Markov decision process (MDP).
 The target of the MU is to choose a set of actions to minimize his cost as shown
 in Eq. (\ref{totalcost}). In this section, we propose a dynamic programming
 based algorithm to solve the problem in Eq. (\ref{totalcost}).\\
 \indent For a MDP problem, it is important to identify the  \textit{optimality equation}
 (or \textit{Bellman equation}) \cite{DPBook}.  Denote $\mathcal{V}_t(\textbf{\textit{s}}_t)$
 as the minimal expected total cost of the MU from $t$ to ${T^M+1}$ at state
 $\textbf{\textit{s}}_t$. The Bellman equation is defined as
 in Eq. (\ref{bellmaneq}).
\begin{equation}\label{bellmaneq}
  \mathcal{V}_t(\textbf{\textit{s}}_t)
  =\min_{\textbf{\textit{a}}_{t}}\Big\{\mathcal{Q}_t(\textbf{\textit{s}}_t,\textbf{\textit{a}}_{t})\Big\}
\end{equation}
where for $l\in\mathcal{L}$, $\textbf{\textit{b}}\in\mathcal{B}$, we have
\begin{equation}\label{bellmaneq2}
\begin{split}
  &\mathcal{Q}_t(\textbf{\textit{s}}_t,\textbf{\textit{a}}_t) 
  =r_t(\textbf{\textit{s}}_t,\textbf{\textit{a}}_t) + \sum_{l_{t+1}\in\mathcal{L}}\sum_{\textbf{\textit{b}}_{t+1}\in\mathcal{B}}\textrm{Pr}(\textbf{\textit{s}}_{t+1}|\textbf{\textit{s}}_t,\textbf{\textit{a}}_t)\mathcal{V}_{t+1}(\textbf{\textit{s}}_{t+1})\\
  =&\underbrace{c_t(\textbf{\textit{s}}_t,\textbf{\textit{a}}_t)+\xi_t(\textbf{\textit{s}}_t,\textbf{\textit{a}}_t)}_{\textrm{cost for the current $t$}}\\
  &+\underbrace{\sum_{l_{t+1}\in\mathcal{L}}\sum_{\textbf{\textit{b}}_{t+1}\in\mathcal{B}}\textrm{Pr}((l_{t+1},\textbf{\textit{b}}_{t+1})|(l,\textbf{\textit{b}}),\textbf{\textit{a}}_t)\mathcal{V}_{t+1}(l_{t+1},\textbf{\textit{b}}_{t+1})}_{\textrm{expected future cost start from $t+1$}}\\
  =&p_c\sum_{j\in\mathcal{M}}\min\{b_t^j,a_{t,c}^j\}\\
  &+\theta_t (\varepsilon_c^l\sum_{j\in\mathcal{M}}\min\{b_t^j,a_{t,c}^j\}+\varepsilon_w^l\sum_{j\in\mathcal{M}}\min\{b_t^j,a_{t,w}^j\})\\
  &+\sum_{l_{t+1}\in\mathcal{L}}p_{l_{t+1},l_t}\mathcal{V}_{t+1}(l_{t+1},[\textbf{\textit{b}}_t-\textbf{\textit{a}}_{t,c}-\textbf{\textit{a}}_{t,w}]^+)\\
\end{split}
\end{equation}
\indent Based on the Bellman equation  Eq. (\ref{bellmaneq}),
we propose Algorithm 1. In the optimal policy calculation
phase, the optimal policy is calculated by backward induction
from epoch $T^M$ to 1, where $\sigma>0$ is the granularity of
the total data size $|\mathcal{B}|$.  Then, the MU's offloading data policy is
decided in each slot in the offloading data transmission phase.
It is obvious that the time complexity of Algorithm 1 is
$\mathcal{O}(|\mathcal{T^M}||\mathcal{L}||\mathcal{B}|/\sigma)$.

\begin{theorem}\label{theorem1}
  The policy $\pi^*=\bigg\{\phi_t^*(l_t,\textbf{\textit{b}}_t),\;\forall\; t\in\mathcal{T}, l\in\mathcal{L}, \textbf{\textit{b}}\in\mathcal{B}\bigg\}$ generated in Algorithm 1 is the problem (\ref{totalcost})'s
  optimal solution.
\end{theorem}
\indent \textit{Proof:} It is obvious according to the principle of optimality defined in \cite{DPBook}.\\
 \begin{flushright}
   \textbf{Q.E.D}
 \end{flushright}

\begin{table}[th]
\renewcommand{\arraystretch}{1}
\begin{tabular}{p{0.1cm}p{7.8cm}}
\hline

\hline
&\textbf{Algorithm 1}: Dynamic Programming Based Algorithm  \\
\hline

\hline
1: &\underline{Optimal Policy Calculation Phase}  \\
2: &Set $\mathcal{V}_{T^M+1}(l,\textbf{\textit{b}})$,$\forall$ $l\in\mathcal{L}$, $\textbf{\textit{k}}\in\mathcal{B}$ by Eq. (\ref{penalty})\\
3: &Set $t$:=$T^M$ \\
4: &\textbf{while} $t\ge1$ : \\
5: &\htab\textbf{for} $l_t \in \mathcal{L}$ : \\
6: &\htab\htab Set $\textbf{\textit{b}}_t:$=0 \\
7: &\htab\htab \textbf{for} $\textbf{\textit{b}}_t \in \mathcal{B}$ :\\
8: &\htab\htab\htab Calculate $\mathcal{Q}_t(\textbf{\textit{s}}_t,\textbf{\textit{a}}_t)$ using Eq. (\ref{bellmaneq})\\
9: &\htab\htab\htab Set $\phi_t^*(l_t,\textbf{\textit{b}}_t)$ := $\arg\min_{\textbf{\textit{a}}_t}\{\mathcal{Q}_t(\textbf{\textit{s}}_t,\textbf{\textit{a}}_t)\}$\\
10: &\htab\htab\htab Set $\mathcal{V}_{t}(l,\textbf{\textit{b}})$ := $\mathcal{Q}_t(\textbf{\textit{s}}_t,\phi_t^*(l_t,\textbf{\textit{b}}_t))$\\
11: &\htab\htab\htab Set $\textbf{\textit{b}}_t$:=$\textbf{\textit{b}}_t+\mathbf{\sigma}$\\
12: &\htab\htab \textbf{end for}\\
13: &\htab \textbf{end for}\\
14: &\htab Set $t$:=$t-1$\\
15: &\textbf{end while} \\
16: &The optimal policy $\mathbf{\pi^*}$ is generated for the following offloading data transmission phase\\
17: & \\
18: & \underline{Offloading Data Transmission Phase}\\
19: & Set $t:=1$, $\textbf{\textit{b}}:=\mathcal{B}$\\
20: & \textbf{while} $t\le T^M$ and $\textbf{\textit{b}}_t > 0$ :\\
21: &\htab $l_t$ is determined from GPS\\
22: &\htab Set action $\textbf{\textit{a}}_t:=\phi_t^*(l_t,\textbf{\textit{b}}_t)$ according to $\mathbf{\pi^*}$ (the optimal policy)\\
25: &\htab\htab Set $\textbf{\textit{b}}_t:=$$[\textbf{\textit{b}}_t-\textbf{\textit{a}}_{t,c}-\textbf{\textit{a}}_{t,w}]^+$ \\
26: &\htab \textbf{end if}\\
27: &\htab Set $t:=t+1$\\
28: & \textbf{end while}\\
\hline

\hline
\end{tabular} \\
\end{table}

\section{Low Time Complexity Heuristic Offloading Algorithm}\label{heuristicalgorithm}
  A dynamic programming based mobile offloading algorithm (Algorithm 1)
  has been proposed in Sect. \ref{dpalgorithm}, and Theorem 1 guarantees
  the optimality of this algorithm. However, time complexity of Algorithm
  1 is rather high. Furthermore, the $\textit{Optimal Policy Calculation Phase}$
  of Algorithm 1 should be performed in advance, which means that Algorithm
  1 is an offline algorithm. Therefore, two questions may arrise.
  \begin{itemize}
  \item \textit{Is there a low time complexity algorithm solution for the MU's problem
  in Eq. (\ref{totalcost})}?
  \item \textit{How to generate the MU's policy in a real-time manner without
  calculations in advance as in Algorithm 1?}
\end{itemize}
\begin{table}[th]
\renewcommand{\arraystretch}{1}
\begin{tabular}{p{0.1cm}p{7.8cm}}
\hline

\hline
&\textbf{Algorithm 2}: Low Time Complexity Heuristic Offloading Algorithm  \\
\hline

\hline
1: &At time slot $t$ \\
2: &\textbf{Input:} $\textbf{\textit{T}}$, $T_{th}$, $l_t$, $\textbf{\textit{b}}_t$\\
3: &\textbf{for} $T^j$$\in$$\textbf{\textit{T}}$:\\
4: &\htab\textbf{if} $t<T^j$:\\
5: &\htab\htab Add $(T^j-t)$ to deadline remain list $\textbf{\textit{R}}$\\
6: &\htab\htab Set $w_t^j$ = $\frac{1}{T^j-t}$ \\
7: &\htab\textbf{else:} \\
8: &\htab\htab Set $w_t^j=0$\\
9: &\htab Add $w_t^j$ to rate allocation weight list $\textbf{\textit{W}}_t$\\
10: &\htab\textbf{end if} \\
11: &\textbf{end for}\\
13: &Normalize $\textbf{\textit{W}}_t$ to $\overline{\textbf{\textit{W}}}_t$\\
13: &Normalize $\textbf{\textit{b}}_t$ to $\overline{\textbf{\textit{b}}}_t$\\
12: &$\textbf{\textit{W}}_t$=multiply($\textbf{\textit{W}}_t,\textbf{\textit{b}}_t$) //multiply by element\\
13: &Normalize $\textbf{\textit{W}}_t$ to $\overline{\textbf{\textit{W}}}_t$\\
14: &\textbf{if } wireless LAN access is available at location $l$ and speed is greater than $\gamma_{th}$:\\
15: &\htab Allocate wireless LAN data rate to each flow according to weight list $\overline{\textbf{\textit{W}}}$.\\
16: &\htab\htab $\textbf{\textit{a}}_{t,w}$ is determined\\
17: &\textbf{else if } $\min (\textbf{\textit{R}}) < T_{th}$:\\
18: &\htab Allocate cellular data rate to each flow according to weight list $\overline{\textbf{\textit{W}}}$.\\
19: &\htab\htab $\textbf{\textit{a}}_{t,c}$ is determined\\
20: &\textbf{end if }\\
21: &\textbf{Output:} $\textbf{\textit{a}}_t = (\textbf{\textit{a}}_{t,c},\textbf{\textit{a}}_{t,w}$)\\
\hline

\hline
\end{tabular} \\
\end{table}
  \indent In this section, 
\ifmark
\textcolor{\encolor}{we try to answer the aforementioned two questions and thus avoid the problems posed by the dynamic programming based Algorithm 1 in Sect. \ref{dpalgorithm}. }
\else
we try to answer the aforementioned two questions and thus avoid the problems posed by the dynamic programming based Algorithm 1 in Sect. \ref{dpalgorithm}.
\fi
  An online low time complexity heuristic offloading algorithm is proposed 
  as shown in Algorithm 2 
\ifmark
\textcolor{\encolor}{by the following arguments:}
\else
  by the following arguments:
\fi  
  \begin{itemize}
  \item (i) a flow with a earlier deadline should have a higher priority;
  \item (ii) the more remaining file size, 
  \ifmark
\textcolor{\encolor}{the higher is the priority of the flow;}
\else
  the higher is the priority of the flow;
\fi  
  \item (iii) the wireless LAN network should have a priority 
  \ifmark
	\textcolor{\encolor}{since it has a lower price than the cellular network;}
  \else
	since it has a lower price than the cellular network;
  \fi  
  \item (iv) when the flow deadline is approaching, the cellular network should
  also be used to try to finish the data transmission, without waiting for 
  access to a wireless LAN network;
  \item (v) a low speed wireless LAN network, which consumes a lot of energy for
  data transmission, should be ignored to 
  \ifmark
	\textcolor{\encolor}{save energy if the MU concerns about the energy consumption;}
  \else
	save energy if the MU concerns about the energy consumption;
  \fi  
  \end{itemize}
\indent We briefly 
  \ifmark
	\textcolor{\encolor}{explain Algorithm 2 below.}
  \else
	explain Algorithm 2 below.
  \fi  

  The main task is 
  \ifmark
	\textcolor{\encolor}{to first calculate the allocation weight $\textbf{\textit{W}}$, }
  \else
	to first calculate the allocation weight $\textbf{\textit{W}}$, 
  \fi 
  \ifmark
	\textcolor{\encolor}{then allocate the wireless LAN data rate or cellular data rate based}
  \else
	then allocate the wireless LAN data rate or cellular data rate based
  \fi 
  on the calculated allocation weight. The inputs of the algorithm 
  are deadline vector $\textbf{\textit{T}}$,
  deadline threshold $T_{th}$ (which will be explained later), location $l$, and
  remaining file size vector $\textbf{\textit{b}}$. The flow with the least remaining
  deadline $({T^j-t})$ has the highest priority, which is calculated as weight
  $w_t^j$ = $\frac{1}{T^j-t}$. Here $T^j$ is the deadline for flow $j\in \mathcal{M}$.
  $\textbf{\textit{W}}_t$ is the weight list for deadlines, which reflects the
  aforementioned rationale (i). 
  \ifmark
	\textcolor{\encolor}{Considering argument (ii) above,}
  \else
	Considering argument (ii) above,
  \fi 
  the weight list for deadlines $\textbf{\textit{W}}_t$ should be 
  multiplied by the remaining
  file size vector $\textbf{\textit{b}}_t$ after normalization
  ($\overline{\textbf{\textit{W}}_t}$ is the normalization of $\textbf{\textit{W}}_t$
  and $\overline{\textbf{\textit{b}}}_t$ is the normalization of $\textbf{\textit{b}}_t$).
  The result of multiplication is denoted as $\textbf{\textit{W}}$. The reason why
  normalization is necessary is that the weight for a deadline and the remaining file
  size are of different scales. While wireless LAN has the priority, we have to use
  wireless LAN when possible. However, if the speed of a wireless LAN is lower than a
  threshold $\gamma_{th}$, the wireless LAN should not be used because it is
  energy-consuming (rationale (v)). Please note that $\gamma_{th}$ is a parameter 
  that is determined by the MU's energy preference $\theta_t$. 
   \ifmark
	\textcolor{\encolor}{If the MU is concerned about energy consumption (high $\theta_t$), }
  \else
	If the MU is concerned about energy consumption (high $\theta_t$), 
  \fi
  the MU will eliminate low speed APs by setting a high 
   threshold $\gamma_{th}$.
  If there is no wireless LAN, the MU has to wait for
  wireless LAN without using cellular network. Yet if the least remaining time for data
  transmission $\min (\textbf{\textit{R}})$ is 
   \ifmark
	\textcolor{\encolor}{less than threshold $T_{th}$, }
  \else
	less than threshold $T_{th}$, 
  \fi
  the cellular network also should be selected for data transmission (rationale (iv)).\\
 \indent It is obvious that the time complexity of Algorithm 2 is $\mathcal{O}(MT^M)$,
 which is much lower than that of Algorithm 1 and there is no offline calculation phase
 for Algorithm 2, therefore, the decision is made in an online manner.
 
\section{Performance Evaluation}\label{performanceevaluation}
In this section, the performances of our dynamic programming
based algorithm (\textit{Proposed DP}) and heuristic offloading
(\textit{Proposed Heuristic}) algorithm are evaluated
by comparing them with a \textit{Baseline} algorithm that use
wireless LAN AP to offload traffic whenever possible and the
algorithm called DAWN in \cite{DAWNHuang2015}.
We developed a simulator with Python 2.7, which can be downloaded
from the following URL link: https://github.com/aqian2006/OffloadingMDP.\\
\indent A four by four grid is used in simulation.
Therefore, $L$ is 16. Wireless LAN APs are randomly
deployed in $L$ locations. The cellular usage price is
assumed as  1.5 yen/Mbyte. $\textrm{Pr}(l|l)=0.6$ means
that the probability that the MU stays in the same place
from time $t$ to $t'$ is 0.6. And the MU moves to the neighbour
location with equal probability, which can be calculated as
$\textrm{p}(l_{t+1}|l)=(1-0.6)/(\textrm{number of neighbors})$.
The average Wireless LAN throughput
$\gamma_{t,w}^l$ is assumed as 15 Mbps\footnote{We tested 
repeatedly with an iPhone 5s on the public wireless LAN APs
of one of the biggest Japanese wireless carriers. 
The average throughput was 15 Mbps. }, while
average cellular network throughput $\gamma_{t,c}^l$ is 10 Mbps\footnote{We
also tested with an iPhone 5s on one of the biggest Japanese wireless
carriers' cellular network. We use the value 10 Mbps for average
cellular throughput.}. 
\ifmark
\textcolor{red}{We generate wireless LAN throughput for each AP
from a truncated normal distribution, and the mean and standard deviation
are assumed as 15Mbps and 6Mbps respectively.  The wireless LAN throughput
is in the range [9Mbps, 21Mbps]. Similarly, we generate cellular throughput 
from a truncated normal distribution, and the mean and standard deviation
are assumed as 10Mbps and 5Mbps respectively.  The cellular network 
throughput is in range [5Mbps, 15Mbps].}
\else
We generate wireless LAN throughput for each AP
from a truncated normal distribution, and the mean and standard deviation
are assumed as 15Mbps and 6Mbps respectively.  The wireless LAN throughput
is in the range [9Mbps, 21Mbps]. Similarly, we generate cellular throughput 
from a truncated normal distribution, and the mean and standard deviation
are assumed as 10Mbps and 5Mbps respectively.  The cellular network 
throughput is in range [5Mbps, 15Mbps].
\fi
$\sigma$ in Algorithm 1 is assumed as 1 Mbits. Time for
each epoch is 1 seconds. The penalty function is assumed as
$g(\textbf{\textit{b}}_t)$=$2\sum_{j\in \mathcal{M}} b_t^j$.
Please refer to Table \ref{tableParm} for the parameters used in
the simulation.\\
\begin{figure}[h]
   \centering
   \includegraphics[width=63mm]{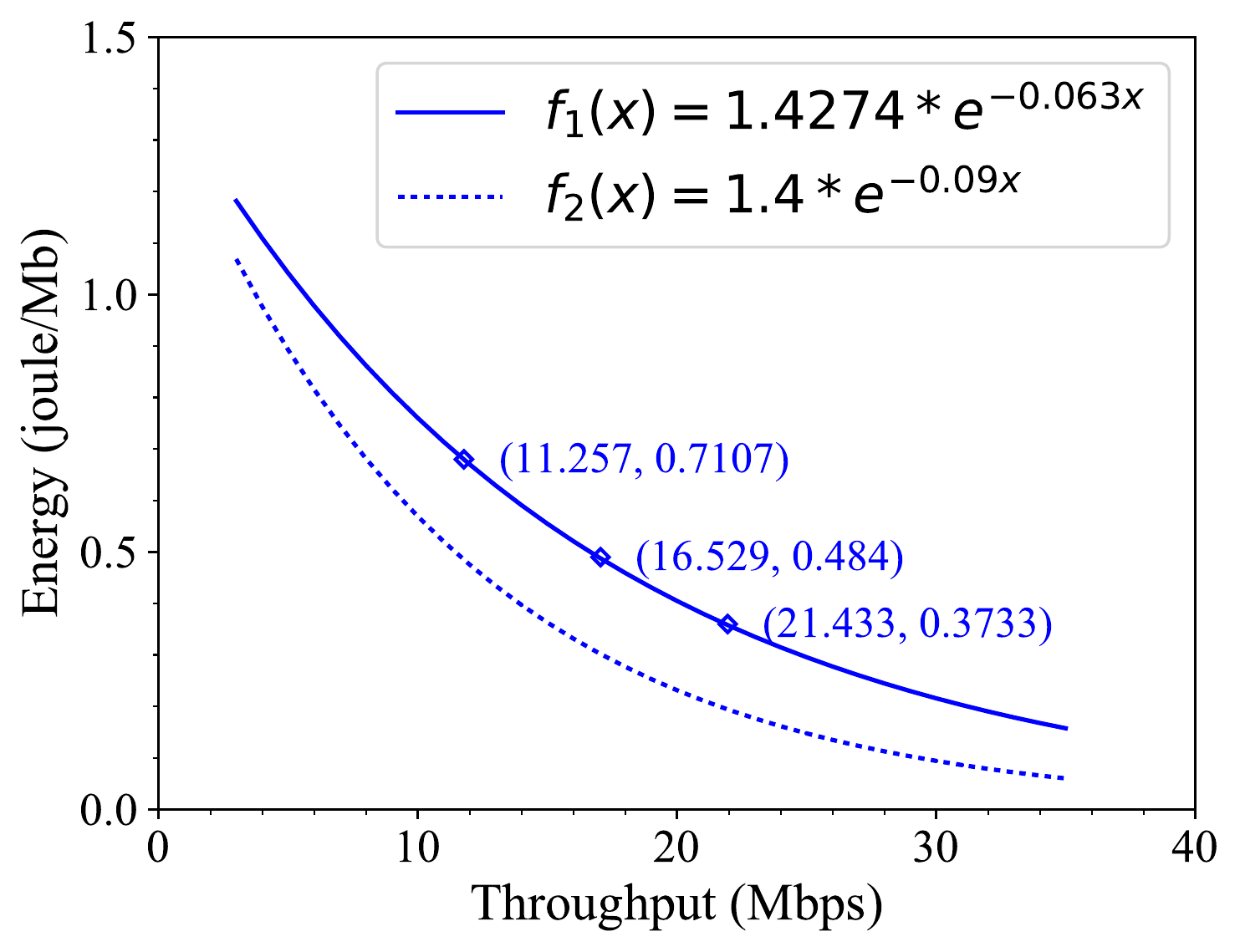}
   \caption{Energy consumption (joule/Mb) vs. Throughput (Mbps).}\label{energythroughput}
\end{figure}
\begin{table}[th]
\caption{Energy vs. Throughput.}
\centering
\label{tableenergy}
\renewcommand{\arraystretch}{1}
\begin{tabular}{|c|c|}
\hline

\hline
Throughput (Mbps) & Energy (joule/Mb)\\
\hline
11.257 &  0.7107\\
\hline
16.529 &  0.484\\
\hline
21.433 &  0.3733\\
\hline

\hline
\end{tabular}
\end{table}
\indent Because the energy consumption rate is a decreasing function of
throughput, we have the sample data from \cite{EnergyThroughput}
(see Table {\ref{tableenergy}). We then fit the sample data by
a exponential function 
\ifmark
\textcolor{blue}{$f_1(x)=1.4274*\mathrm{e}^{-0.063x}$}
\else
$f_1(x)=1.4274*\mathrm{e}^{-0.063x}$
\fi
as shown in Fig.  \ref{energythroughput}.
\ifmark
\textcolor{red}{We also made a new energy-throughput
function as $f_2(x)=1.4*\mathrm{e}^{-0.09x}$, which is just lower than
$f_1(x)$. We basically use $f_1(x)$ if we do not explicitly point out.}
\else
We also made a new energy-throughput
function as $f_2(x)=1.4*\mathrm{e}^{-0.09x}$, which is just lower than
$f_1(x)$.
We basically use $f_1(x)$ if we do not explicitly point out.
\fi
Please note that the energy consumption rate of cellular and wireless LAN
may be different for the same throughput, but we assume
they are the same and use the same fitting function as in Fig. \ref{energythroughput}.
\iftable
\begin{table}[t]
\caption{Parameters in the simulation.}\label{tableParm}
\centering
\label{tableparameters}
\renewcommand{\arraystretch}{1}
\begin{tabular}{|c|c|}
\hline

\hline
Parameters & $\hspace{1cm}$Value$\hspace{1cm}$\\
\hline

\hline
 $L$ &16\\
\hline
 $\textbf{\textit{B}}$ &$\textbf{\textit{B}}=(500,550,600,650)$ Mbits\\
\hline
 $\mathcal{T}$ &$\mathcal{T}=(140,280,420,560)$\\
\hline
Number of wireless LAN APs &8\\
\hline
$\sigma$ & 1 Mbits\\
\hline
time slot & 1 seconds\\
\hline
average of $\gamma_{c}^l$ & 10 Mbps\\
\hline
\ifmark
\textcolor{red}{standard deviation of $\gamma_{c}^l$} & \textcolor{red}{5 Mpbs}\\
\else
standard deviation of $\gamma_{c}^l$ & 5 Mpbs\\
\fi
\hline
average of $\gamma_{w}^l$ & 15 Mbps\\
\hline
\ifmark
\textcolor{red}{standard deviation of $\gamma_{w}^l$} & \textcolor{red}{6 Mpbs}\\
\else
standard deviation of $\gamma_{w}^l$ & 6 Mpbs\\
\fi
\hline
$p_{l,l}$ & 0.6\\
\hline
$p_{l_{t+1},l_t}$ & (1-0.6)/\#neigbour locations\\
\hline
\ifmark
$p_c$ & \textcolor{blue}{1.5 yen per Mbyte}\\
\else
$p_c$ & 1.5 yen per Mbyte\\
\fi
\hline
$g(\textbf{\textit{b}}_t)$ & $g(\textbf{\textit{b}}_t)$=$2\sum_{j\in \mathcal{M}} b_t^j$\\
\hline

\hline
\end{tabular}
\end{table}
\else
\fi
\begin{figure}[h]
   \centering
   \includegraphics[width=\figwidth]{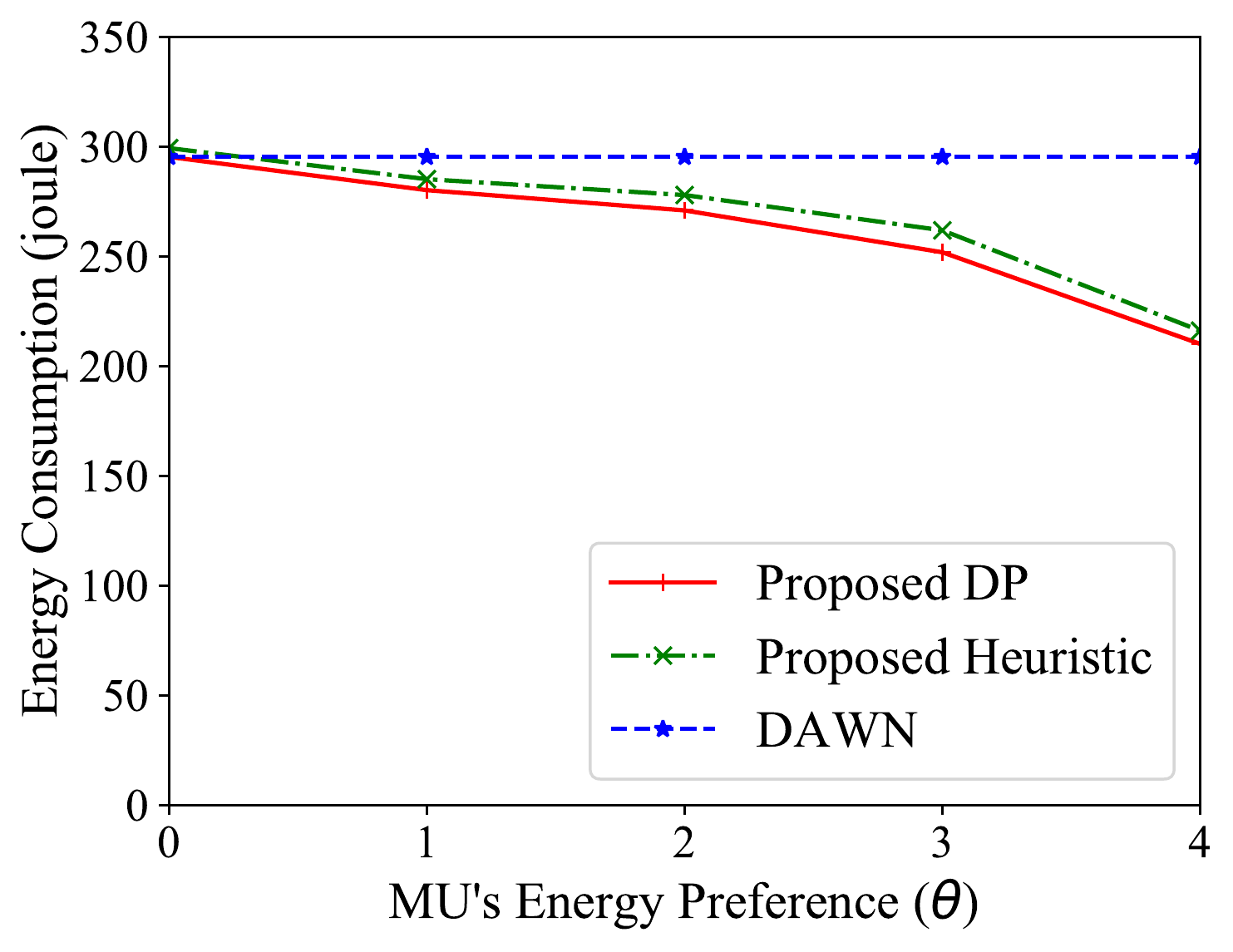}
   \caption{Energy consumption (joule) vs. MU's energy preference ($\theta$).}
   \label{FigEnergyCompPref}
\end{figure}

\indent In Fig.\ref{FigEnergyCompPref}, our proposed DP and heuristic algorithms are compared to the DAWN algorithm in \cite{DAWNHuang2015} in terms
of the MU's energy consumption. Since \cite{DAWNHuang2015} only considered 
a single-flow case, we also apply our algorithms to a single-flow. It is shown
that the larger the MU's energy preference, the lower the energy consumed for our
MDP and heuristic algorithms. The energy consumption of the DAWN algorithm 
is higher than that of our algorithm. The heuristic algorithm is not optimal,
but it is close  to the optimal result of proposed DP algorithm.
\ifmark
\textcolor{red}{The reasons is that energy consumption was not considered in
the DAWN algorithm, while our proposed DP and heuristic algorithms have taken energy 
consumption into consideration and tried to minimize total energy consumption. }
\else
The reasons is that energy consumption was not considered in
the DAWN algorithm, while our proposed DP and heuristic algorithms have taken energy 
consumption into consideration and tried to minimize total energy consumption.
\fi

\begin{figure}[h]
   \centering
   \includegraphics[width=\figwidth]{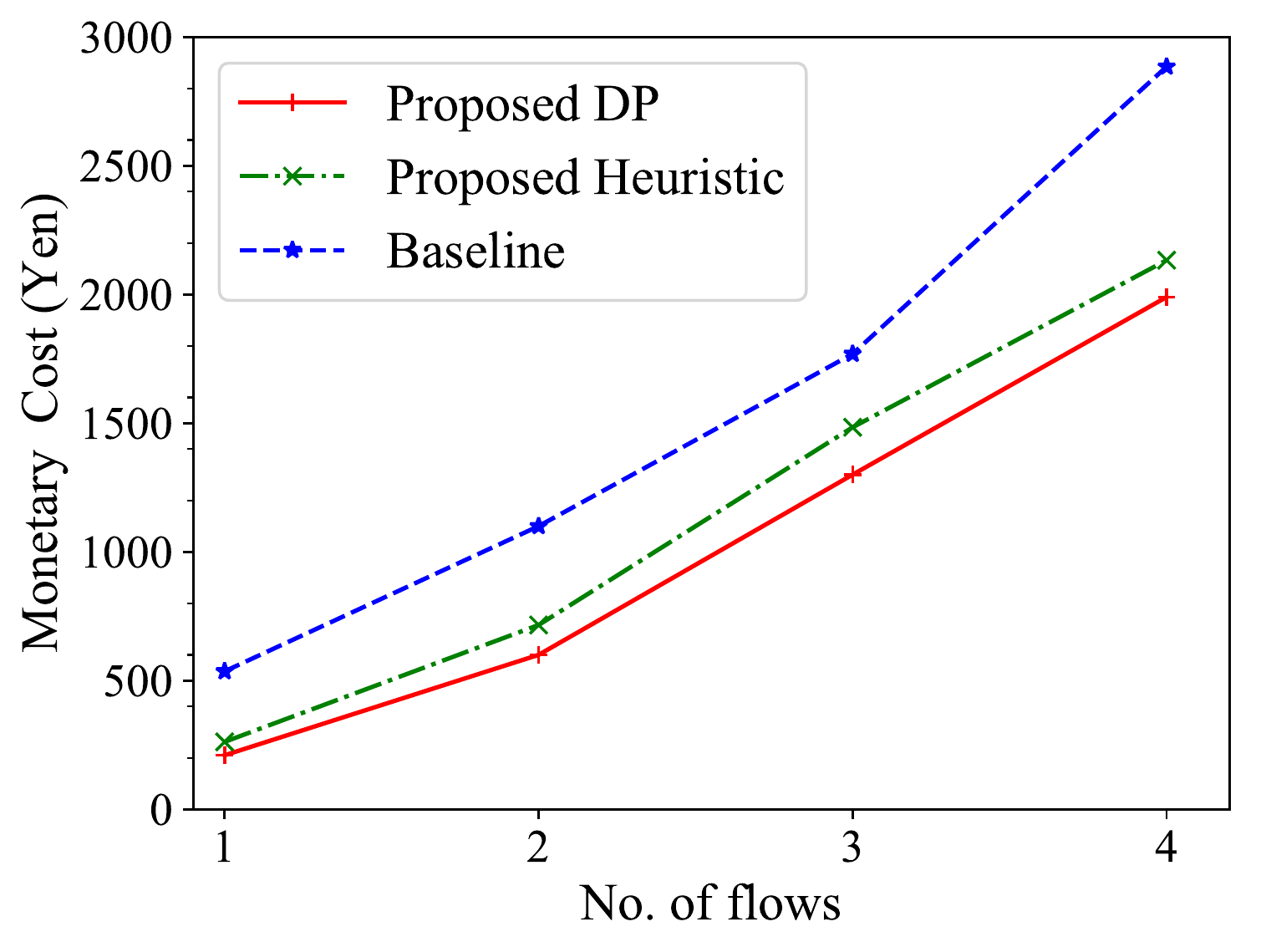}
   \caption{Monetary cost (yen) vs. No. of flows.}\label{FigCostFlows}
\end{figure}

\begin{figure}[h]
   \centering
   \includegraphics[width=\figwidth]{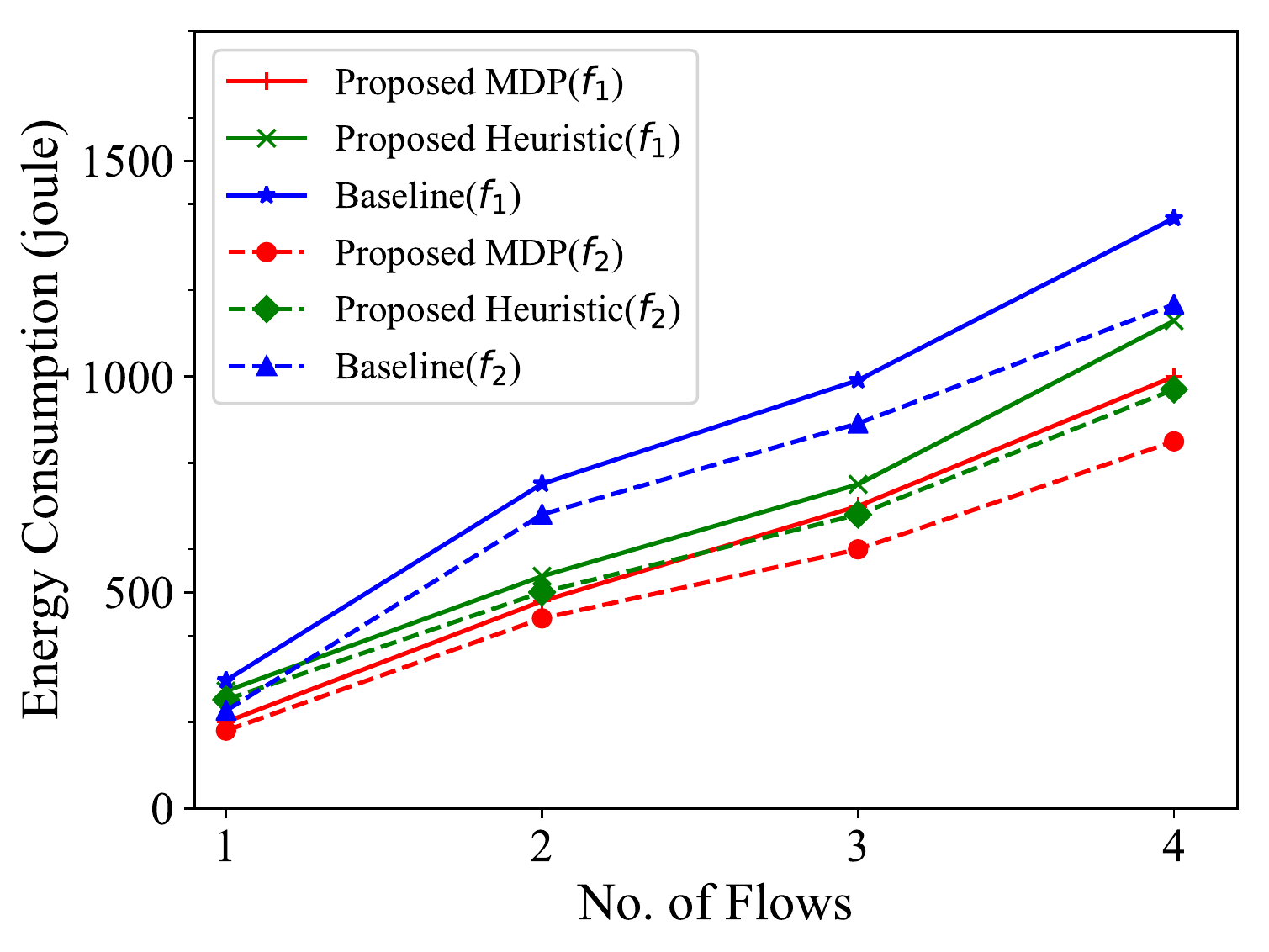}
\ifmark
\textcolor{blue}{\caption{Energy consumption (joule) vs. No. of flows with different energy-throughput functions $f_1$ and $f_2$.}\label{FigEnergyCompFLows}}
\else
\caption{Energy consumption (joule) vs. No. of flows with different energy-throughput functions $f_1$ and $f_2$.}\label{FigEnergyCompFLows}
\fi
\end{figure}

\ifmark
\textcolor{red}{Fig.\ref{FigCostFlows} shows the comparison of monetary cost among \textit{Baseline}, \textit{Proposed Heuristic}, and \textit{Proposed DP} algorithms with different 
number of flows.}
\else
Fig.\ref{FigCostFlows} and shows the comparison of monetary cost among \textit{Baseline}, \textit{Proposed Heuristic}, and \textit{Proposed DP} algorithms with different 
number of flows.
\fi
\ifmark
\textcolor{red}{The monetary cost of all three algorithms increases with the number of flows.
The monetary cost of \textit{Proposed DP} is lower than \textit{Baseline}, while 
\textit{Proposed Heuristic} is close to \textit{Proposed DP}. The reason is that 
in \textit{Baseline}, data are downloaded whenever there is a network (cellular or wireless LAN) available, without considering the monetary cost by using the cellular network.
In \textit{Proposed DP} and \textit{Proposed Heuristic}, whether the cellular network 
is used depends on the remaining data to download and the deadline. If there are only
relatively few remaining data and enough time left until the deadline, our proposed algorithms will choose to wait for a cheap wireless LAN to download data.}
\else
The monetary cost of all three algorithms increases with the number of flows.
The monetary cost of \textit{Proposed DP} is lower than \textit{Baseline}, while 
\textit{Proposed Heuristic} is close to \textit{Proposed DP}. The reason is that 
in \textit{Baseline}, data are downloaded whenever there is a network (cellular or wireless LAN) available, without considering the monetary cost by using the cellular network.
In \textit{Proposed DP} and \textit{Proposed Heuristic}, whether the cellular network 
is used depends on the remaining data to download and the deadline. If there are only
relatively few remaining data and enough time left until the deadline, our proposed algorithms will choose to wait for a cheap wireless LAN to download data.
\fi

Fig.\ref{FigEnergyCompFLows} shows the comparison of the energy consumption among \textit{Baseline}, \textit{Proposed Heuristic}, and \textit{Proposed DP} algorithms with different number of flows.  Two energy-throughput functions $f_1(x)$ and $f_2(x)$ 
are used. The performance of \textit{Proposed DP} algorithm is the best, but the \textit{Proposed Heuristic} algorithm shows small differences with that of \textit{Proposed DP} algorithm with either $f_1(x)$ or $f_2(x)$. 
\ifmark
\textcolor{red}{For \textit{Proposed DP}/\textit{Proposed Heuristic}/\textit{Baseline}, 
the energy consumption under $f_1(x)$ is much higher than that under $f_2(x)$. 
The reason is that the energy consumption for a certain throughput is higher under 
$f_1(x)$ than that under $f_2(x)$. Our \textit{Proposed DP} algorithm consumes 
the least energy since we attempt to minimize the total energy consumption by formulating 
a MDP problem.}
\else
For \textit{Proposed DP}/\textit{Proposed Heuristic}/\textit{Baseline}, 
the energy consumption under $f_1(x)$ is much higher than that under $f_2(x)$. 
The reason is that the energy consumption for a certain throughput is higher under 
$f_1(x)$ than that under $f_2(x)$. Our \textit{Proposed DP} algorithm consumes 
the least energy since we attempt to minimize the total energy consumption by formulating 
a MDP problem.
\fi

\begin{figure}[h]
   \centering
   \includegraphics[width=\figwidth]{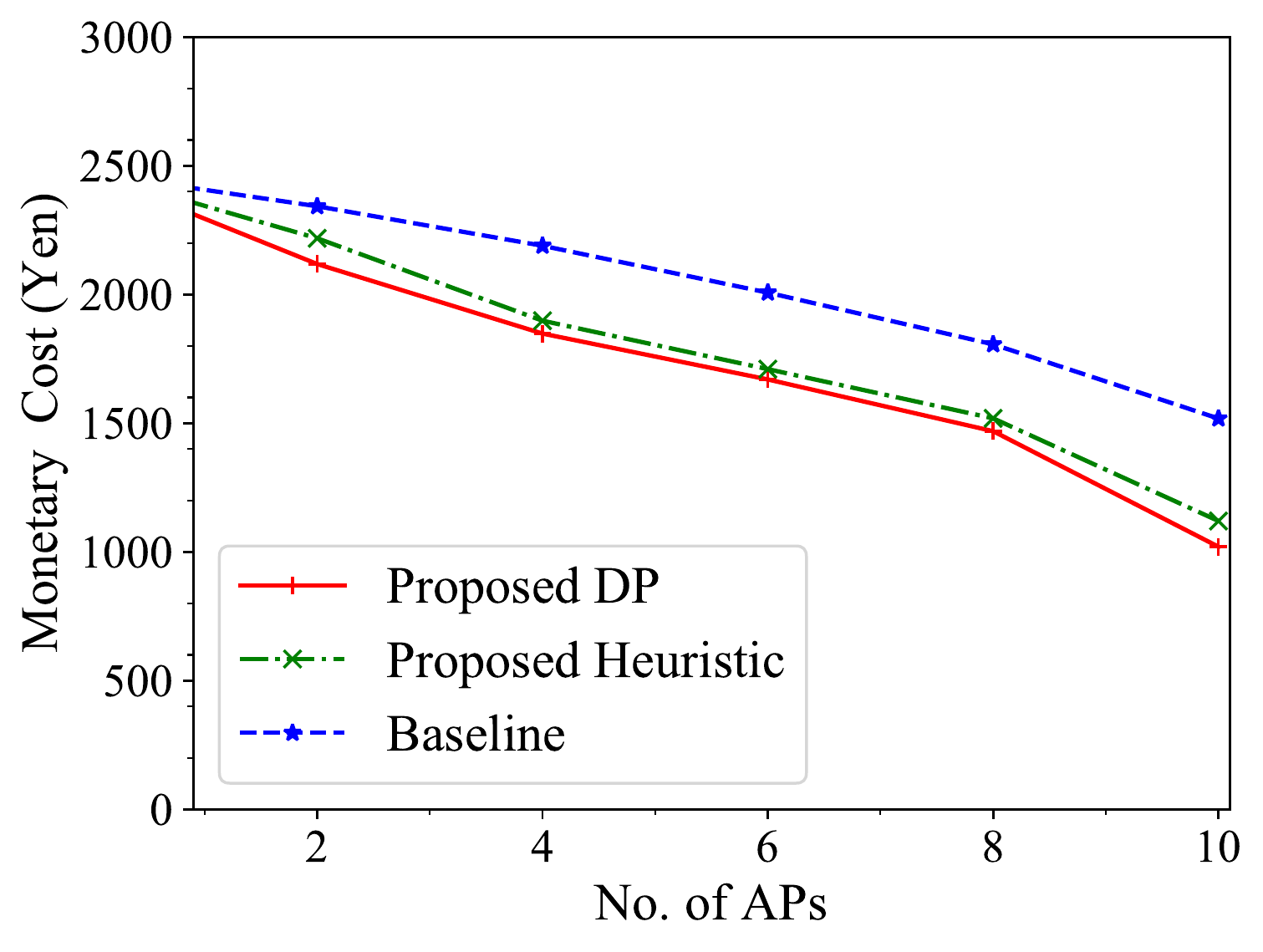}
   \caption{Monetary cost (yen) vs. No. of APs.}\label{FigCostAPs}
\end{figure}

\begin{figure}[h]
   \centering
   \includegraphics[width=\figwidth]{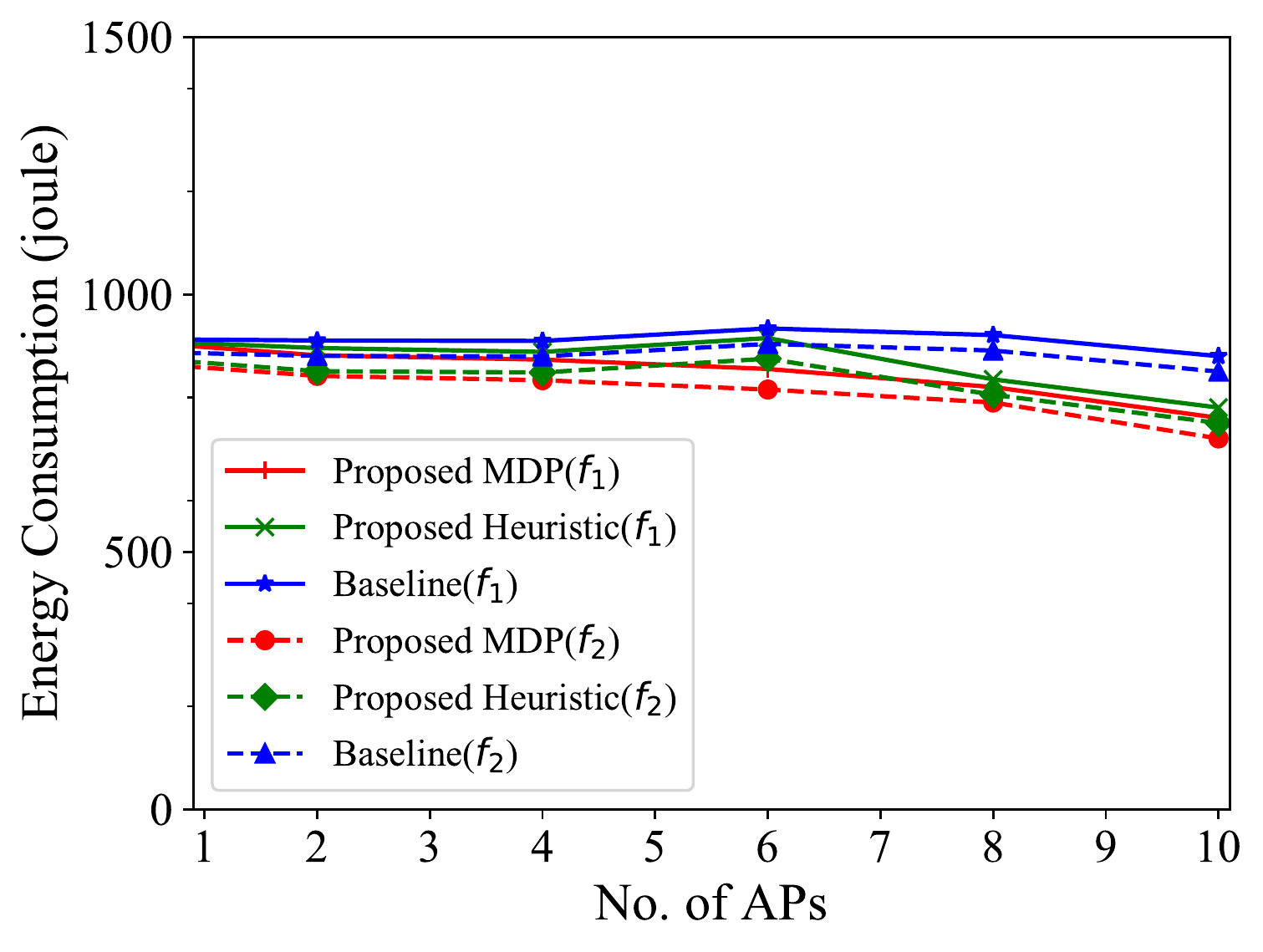}
\ifmark
\textcolor{blue}{\caption{Energy consumption (joule) vs. No. of APs with different energy-throughput functions $f_1$ and $f_2$.}\label{FigEnergyAPs}}
\else
\caption{Energy consumption (joule) vs. No. of APs with different energy-throughput functions $f_1$ and $f_2$.}\label{FigEnergyAPs}
\fi
\end{figure}

\begin{figure}[h]
   \centering
   \includegraphics[width=\figwidth]{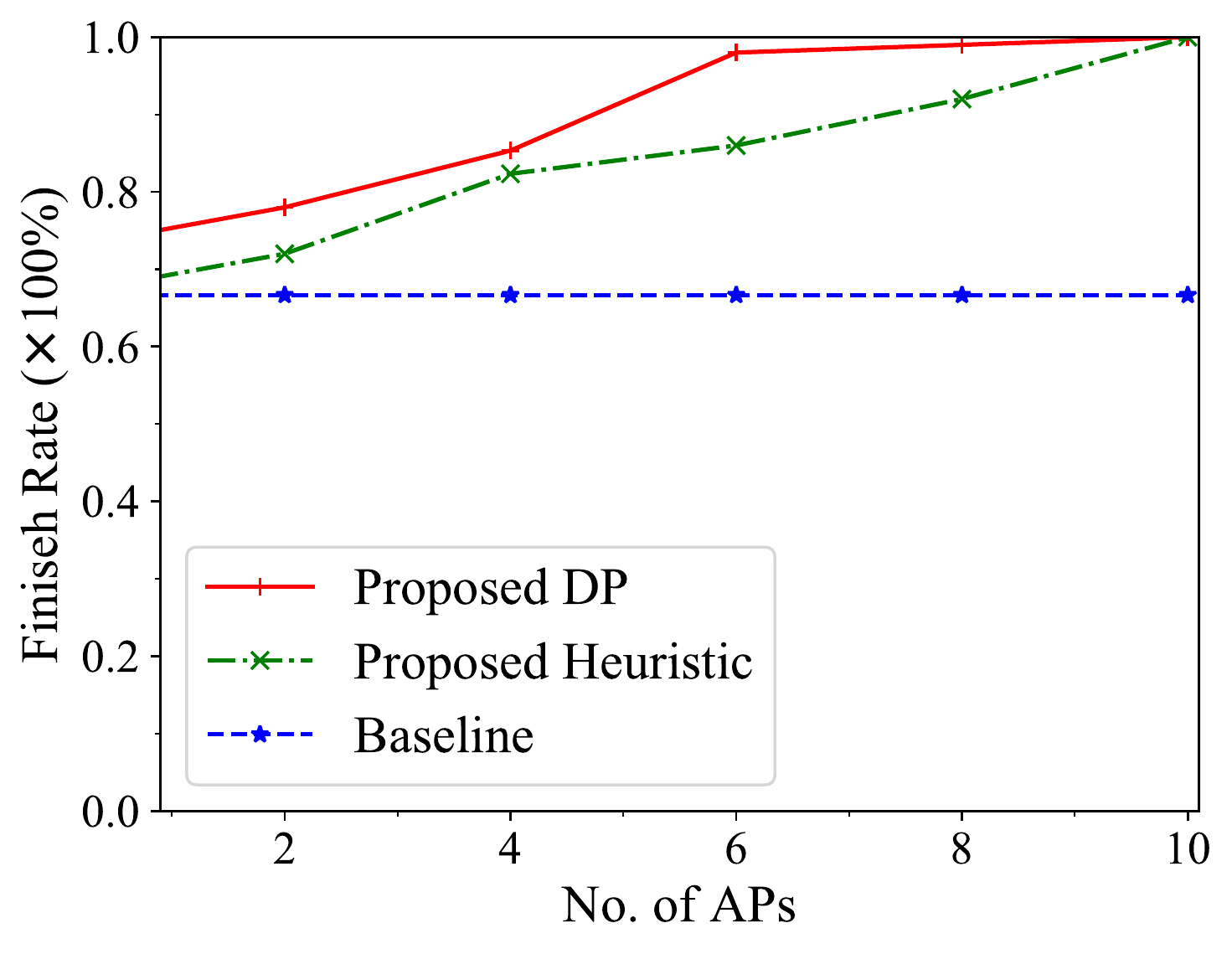}
   \caption{Finish rate vs. No. of APs.}\label{FigFinishrateAPs}
\end{figure}

\ifmark
Fig.\ref{FigCostAPs} shows the comparison of monetary cost among \textit{Baseline}, \textit{Proposed Heuristic}, and \textit{Proposed DP} algorithms \textcolor{blue}{with different 
number of APs.}
\else
Fig.\ref{FigCostAPs} shows the comparison of monetary cost among \textit{Baseline}, \textit{Proposed Heuristic}, and \textit{Proposed DP} algorithms with different 
number of APs.
\fi
\ifmark
\textcolor{blue}{It can be seen that the monetary cost of \textit{Proposed DP} algorithm is lowest, and the baseline algorithm is highest.} 
\else
It can be seen that the monetary cost of \textit{Proposed DP} algorithm is lowest, and the baseline algorithm is highest.
\fi
While the monetary cost of the \textit{Proposed Hueristic} algorithm is between 
that of \textit{Baseline}, it is much closer to the \textit{Proposed DP} algorithm.
\ifmark
\textcolor{red}{With a large number of wireless LAN APs deployed, the chance 
of using cheap wireless LAN increases. Therefore, the MU can reduce his monetary 
cost by using cheap wireless LAN. Therefore, all three algorithms' monetary 
costs decreases with the number of APs.}\\
}
\else
With a large number of wireless LAN APs deployed, the chance 
of using cheap wireless LAN increases. Therefore, the MU can reduce his monetary 
cost by using cheap wireless LAN. Therefore, all three algorithms' monetary 
costs decreases with the number of APs.\\
\fi
\indent Fig.\ref{FigEnergyAPs} shows how the MU's energy consumption changes with
the number of deployed APs under the two energy-throughput functions $f_1(x)$ and $f_2(x)$.
\ifmark
\textcolor{red}{Similar to Fig.\ref{FigEnergyCompFLows}, the performance of \textit{Proposed DP} algorithm is the best with either $f_1(x)$ or $f_2(x)$.  
It shows that the energy consumptions of all three algorithms just  slightly decrease
with the number of APs. The reason is that the energy consumption depends
on the throughput. The larger the throughput, the lower is the energy consumption.
With large number of wireless LAN APs, the MU has more chance to use wireless LAN 
with high throughput since the average throughput of a wireless LAN is assumed 
as higher than that of cellular network (see Table \ref{tableParm}). }\\
\else
Similar to Fig.\ref{FigEnergyCompFLows}, the performance of \textit{Proposed DP} algorithm is the best with either $f_1(x)$ or $f_2(x)$.  
It shows that the energy consumptions of all three algorithms just  slightly decrease
with the number of APs. The reason is that the energy consumption depends
on the throughput. The larger the throughput, the lower is the energy consumption.
With large number of wireless LAN APs, the MU has more chance to use wireless LAN 
with high throughput since the average throughput of a wireless LAN is assumed 
as higher than that of cellular network (see Table \ref{tableParm}). \\
\fi
\indent Fig.\ref{FigFinishrateAPs} shows the \textit{finish rate} comparison among \textit{Basedline}, \textit{Proposed DP} and the \textit{Proposed Heuristic} algorithm with different number of wireless LAN APs. 
\ifmark
\textcolor{red}{Here, finish rate is defined as the ratio of the number of flows with finished transmission to the total number of flows started. Even though there are penalties 
for flows' remaining data, not all the flows can be finished before their deadlines.}
\else
Here, finish rate is defined as the ratio of the number of flows with finished transmission to the total number of flows started. Even though there are penalties 
for flows' remaining data, not all the flows can be finished before their deadlines.
\fi
Finish rate of \textit{Proposed DP} and the \textit{Proposed Heuristic} algorithms increases with the number of wireless LAN APs deployed.
\ifmark
\textcolor{red}{The reason is that a large number of  cheap and high throughput
wireless LAN APs decreases the overall download time.}
\else
The reason is that a large number of  cheap and high throughput
wireless LAN APs decreases the overall download time.
\fi

\ifmemo
\begin{table}[th]
\textcolor{red}{\caption{Figures of simulation section.}}
\centering
\label{simulatioin-todo}
\renewcommand{\arraystretch}{1}
\begin{tabular}{|c|c|c|c|}
\hline

\hline
\textcolor{red}{No.}& x-axis & y-axis & parameter\\
\hline
\textcolor{red}{1}& energy preference ($\theta$)& energy consumption & \#flows, AP coverage\\
\hline
\textcolor{red}{2} & \#flow & monetary cost & $\theta$, AP coverage\\
\hline
\textcolor{red}{3} & \#flow & energy consumption & $\theta$, AP coverage\\
\hline
\textcolor{red}{4} & AP coverage & monetary cost & \#flow, $\theta$\\
\hline
\textcolor{red}{5} & AP coverage & energy consumption & \#flow, $\theta$\\
\hline
\textcolor{red}{6} & deadline & monetary cost & \#flow, $\theta$, AP coverage\\
\hline
\textcolor{red}{7} & deadline & energy consumption & \#flow, $\theta$, AP coverage\\
\hline
\textcolor{red}{8} & deadline & prob. completion & \#flow, $\theta$, AP coverage\\

\hline

\hline

\end{tabular}
\end{table}
\else
\fi
\ifmark
\indent \textcolor{red}{There are two limitations for our proposed DP and heuristic algorithms: (i) the proposed DP algorithm has a very high time-complexity, therefore it 
takes time to get the optimal policy for the MU.  Therefore, we proposed a low time-complexity heuristic algorithm for the MU.
(ii) The heuristic algorithm can compute the policy very fast, and the simulation results 
have shown that the performance is comparable with our optimal DP algorithm. But we have not theoretically proofed yet that the heuristic algorithm is optimal or near optimal.}\\
\else
There are two limitations for our proposed DP and heuristic algorithms: (i) the proposed DP algorithm has a very high time-complexity, therefore it 
takes time to get the optimal policy for the MU.  Therefore, we proposed a low time-complexity heuristic algorithm for the MU.
(ii) The heuristic algorithm can compute the policy very fast, and the simulation results 
have shown that the performance is comparable with our optimal DP algorithm. But we have not theoretically proofed yet that the heuristic algorithm is optimal or near optimal.
\fi

\section{Conclusion}\label{conclusion}
In this paper, we studied a multi-flow mobile data offloading problem 
\ifmark
\textcolor{\encolor}{in which a MU has multiple applications that want to download data simultaneously}
\else
in which a MU has multiple applications that want to download data simultaneously
\fi
with different deadlines. We formulated the wireless LAN
offloading problem as a finite-horizon discrete-time Markov decision
process.\\
\indent  A dynamic programming based offloading algorithm was proposed 
\ifmark
\textcolor{\encolor}{and its time complexity was analyzed.}
\else
and its time complexity was analyzed.
\fi
\ifmark
\textcolor{\encolor}{Analysis results showed that the}
\else
Analysis results showed that the 
\fi
time complexity of the algorithm is rather high. We proposed a 
low time complexity heuristic offloading algorithm.
Extensive simulations have shown that the DP algorithm
had the lowest cost, while the heuristic algorithm had comparable
performance as that of DP algorithm.\\
\ifmark
\indent \textcolor{red}{This work assumes that the MNO adopts 
usage-based pricing, in which the MU paid for the MNO in proportion with data usage.
In the future, we will evaluate other usage-based pricing
variants like tiered data plan, in which the payment of the MU is 
a step function of data usage. And we will also use time-dependent pricing
(TDP) we proposed in  \cite{TimeDependentPriceDuopolyNSP}\cite{TimeDependentPriceOligopolyNSP}, without changing 
the framework and algorithms proposed in this paper.}
\else
This work assumes that the MNO adopts 
usage-based pricing, in which the MU paid for the MNO in proportion with data usage.
In the future, we will evaluate other usage-based pricing
variants like tiered data plan, in which the payment of the MU is 
a step function of data usage. And we will also use time-dependent pricing
(TDP) we proposed in  \cite{TimeDependentPriceDuopolyNSP}\cite{TimeDependentPriceOligopolyNSP}, without changing 
the framework and algorithms proposed in this paper.
\fi
We also assume that the MU can only use one network interface at most,
either cellular network or wireless LAN, at each time. In 
future work, we will relax this assumption to see 
how the energy consumption and monetary cost will be in this case.
Another assumption we have made is that the MU's mobile probability 
from one place to another is known. It is reasonable if the MU moves 
in a certain pattern, for example, people may commute from home
to work by the same train at the same time in weekdays. In our 
future work, we would also like to consider the case 
\ifmark
\textcolor{\encolor}{wherein the probability of the the MU's movement from one place to another is unknown.}
\else
wherein the probability of the MU's movement from one place to another is unknown.
\fi
The Possible solution is to utilize learning technology to predict the MU's mobile probability.

\section*{Acknowledgements}\label{Acknowledgements}
This work is part of the Grant-in-Aid for Young Scientists (B)
research programme with grant number 16K18109, which is
financed by the Japan Society for the Promotion of Science (JSPS).

\bibliographystyle{IEEEtran}
\bibliography{reference}
\ifCLASSOPTIONcaptionsoff
  \newpage
\fi

\end{document}